# Benefits and Cyber-Vulnerability of Demand Response System in Real-Time Grid Operations

Mingjian Tuo
*Student Member, IEEE*
Department of Electrical and Computer Engineering
University of Houston
Houston, TX, USA
mtuo@uh.edu

Arun Venkatesh Ramesh
*Student Member, IEEE*
Department of Electrical and Computer Engineering
University of Houston
Houston, TX, USA
aramesh4@uh.edu

Xingpeng Li
*Member, IEEE*
Department of Electrical and Computer Engineering
University of Houston
Houston, TX, USA
xli82@uh.edu

*Abstract*— With improvement in smart grids through two-way communication, demand response (DR) has gained significant attention due to the inherent flexibility provided by shifting non-critical loads from peak periods to off-peak periods, which can greatly improve grid reliability and reduce cost of energy. Operators utilize DR to enhance operational flexibility and alleviate network congestion. However, the intelligent two-way communication is susceptible to cyber-attacks. This paper studies the benefits of DR in security-constrained economic dispatch (SCED) and then the vulnerability of the system to line overloads when cyber-attack targets DR signals. This paper proposes a false demand response signal and load measurement injection (FSMI) cyber-attack model that sends erroneous DR signals while hacking measurements to make the attack undetectable. Simulation results on the IEEE 24-bus system (i) demonstrate the cost-saving benefits of demand response, and (ii) show significant line overloads when the demand response signals are altered under an FSMI attack.

*Index Terms*— Controllable load curtailment, Cyber-attack, Demand response, Demand response signal redistribution, False data injection, False load measurement injection, Linear programming, Optimization, Security-constrained economic dispatch.

## NOMENCLATURE

*Sets:*
| | |
|---|---|
| $G$ | Set of online generators. |
| $G(n)$ | Set of online generators connected to bus $n$. |
| $K$ | Set of lines. |
| $K^+(n)$ | Set of lines with bus $n$ as receiving bus. |
| $K^-(n)$ | Set of lines with bus $n$ as sending bus. |
| $T$ | Set of time periods. |
| $N$ | Set of buses. |

*Indices:*
| | |
|---|---|
| $g$ | Generator $g$. |
| $k$ | Line $k$. |
| $t$ | Time $t$. |
| $n$ | bus $n$. |
| $l$ | Targeted attack line for line over-load |

*Parameters:*
| | |
|---|---|
| $c_g$ | Linear operation cost for generator $g$. |
| $c_{DR}^{15}$ | Cost for shifted demand response by 1 period. |
| $c_{DR}^{30}$ | Cost for shifted demand response by 2 periods. |
| $c_{DR}^{45}$ | Cost for shifted demand response by 3 periods. |
| $P_g^{min}$ | Minimum output limit of generator $g$. |
| $P_g^{max}$ | Maximum output limit of generator $g$. |
| $P_k^{max}$ | Long-term thermal line limit for line $k$. |
| $b_k$ | Susceptance of line $k$. |
| $d_{n,t}$ | Forecasted demand at bus $n$ in time period $t$. |
| $\alpha$ | Load shift deviation factor. |
| $DR_n^{Max}$ | Maximum DR participation at bus $n$. |
| $\widetilde{DR}_n^{Max}$ | Maximum post-attack DR under at bus $n$. |
| $\widetilde{DR}_n^{Min}$ | Minimum post-attack DR under at bus $n$. |
| $S_0$ | Limit of cumulative phase angle deviation $l_1$-norm constraint. |
| $Q_0$ | Limit of total DR deviation $l_1$-norm constraint. |
| $netDmnd_{n,t}$ | Net-demand served at time $t$ at node $n$. |
| $netDmnd_n^{max}$ | Maximum net-demand served at node $n$. |
| $\Delta T$ | Short-term 15-minute period/time window. |

*Variables:*
| | |
|---|---|
| $P_{g,t}$ | Output of generator $g$ in time period $t$. |
| $c_{n,1}$ | Deviation in phase angle of bus $n$. |
| $q_n$ | Deviation in shifted demand at bus $n$. |
| $DR_{n,t}^{15}$ | Demand shifted by 1 period at bus $n$ in time period $t$. |
| $DR_{n,t}^{30}$ | Demand shifted by 2 periods at bus $n$ in time period $t$. |
| $DR_{n,t}^{45}$ | Demand shifted by 3 periods at bus $n$ in time period $t$. |
| $DR_{n,1}$ | Total scheduled demand shifted at bus $n$ in period 1. |
| $\widetilde{DR}_{n,1}$ | False demand scheduled to be shifted at bus $n$ in period 1. |
| $\Delta\widetilde{DR}_{n,1}$ | DR deviation at bus $n$ in time period 1. |
| $P_{k,t}$ | Flow on line $k$ in time period $t$. |
| $\tilde{P}_{k,1}$ | Attacked flow on line $k$ in first time period. |
| $\theta_{n,t}$ | Phase angle of bus $n$ in time period $t$. |
| $\theta_{m,t}$ | Phase angle of bus $m$ in time period $t$. |
| $\tilde{\theta}_{n,t}$ | Post- attack phase angle of bus $n$ in period $t$. |
| $\tilde{\theta}_{m,t}$ | Post- attack phase angle of bus $m$ in period $t$. |

## I. INTRODUCTION

The advancement in smart grid technologies has brought two-way communications along with sensing and control signals. This requires an intelligent energy management system (EMS) and an enhanced cyber-layer to efficiently and securely operate. These technologies enable system operators to determine and send signals to redispatch generators and enable demand response for controllable loads. However, this increases the vulnerability of the system to cyber-attacks. The key components of EMS include supervisory control and data acquisition (SCADA), state estimation with bad-data detection, real-time contingency analysis (RTCA) and security-constrained economic dispatch (SCED) [1], as shown in Fig. 1 (a).

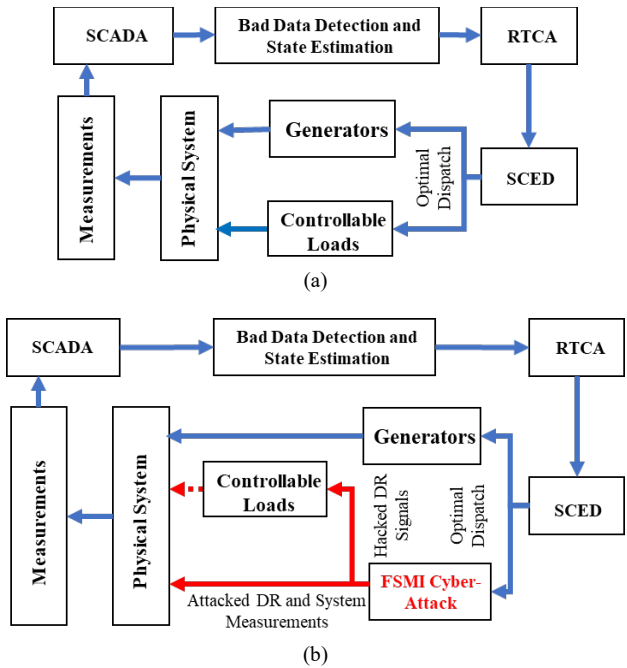

Fig. 1. Illustration of power grid real-time operations (a) under normal operation (b) under the proposed FSMI attack.

Here, SCADA consolidates all the measurements, and then the state estimation along with the bad data detection can work iteratively to estimate the true system state variables. The RTCA process is a sequence of power flow runs using the state estimation solution to identify critical contingencies in the system. The SCED determines the physical signals for least-cost generator dispatch solution to meet the forecasted demand, the information from RTCA is used in SCED to ensure reliability and security of the system [2].

The SCED is not only used to determine the least-cost generator dispatch solution but also obtain demand response (DR) signals. This problem is solved in real-time and short-term operations in both regulated and deregulated environments. DR through controllable loads benefits the system by moving non-critical deferrable loads from peak hours to non-peak hours which increases the system flexibility [3]. Therefore, the operators require DR participating loads to be dispatchable in the SCED. The loads participating in DR programs are compensated through capacity markets and/or real-time markets [4]-[9].

Before the automated DR was implemented, load controllability was implemented through manual intervention which had lower chances of cyber-attacks. However, with the increasing use of standard products, operating systems and distributed networking devices add intelligence and two-way communication capabilities to the smart grid [10], which results in automated demand response. This eventually means another access for cyber vulnerability attacks to the system.

There is an extensive interest in the vulnerability of power systems and cyber-attacks on various elements of the system such as attacks on system state estimation [11], system topology [12], generator dynamics [13], and energy markets [14]. A popular cyber-attack targets overloading the system, i.e. overloading one or multiple lines, which may cause large system disturbances and lead to cascade events [15]-[16]. In Fig. 1 (b), it is perceived that sending erroneous DR signals along with coordinated false measurements can be undetectable to the system operator.

In markets that offer DR services, consumers have the option to register for DR programs that are run by utilities. The utility offers incentives to consumers to either lower or shift their electric demand to improve power system balancing and reduce the overall system cost. Utilities use a client device referred to as a DR client to communicate with the utility DR server [17]. Such technologies resulted in an increase in DR programs participation in energy markets to 29,674 MW in 2018, 7.7% higher than 2017 [18]. When the loads are required to shift, the DR server will send commands to the DR clients, then the DR clients either defer or drop the loads following the DR signal from the server.

Such timely communication requires the availability of the communication channels and server systems especially for fast-DR to control the electricity demand within a very short period. Not only that, the integrity of data requires the curtailment signal should only be sent by operator and only participating customers receive DR signals. The availability of the communication network and the integrity of DR data are essential for information security. Without these standard information securities, signals can be maliciously manipulated to pose significant impact to the system. Therefore, with such vulnerabilities, the automated DR system that brings great performance benefit to the grid can also pose great risks and challenges in protecting the smart grids from cyber security threats [19]. [20] investigated an automated DR process and proposed a security-enhancement mechanism to enhance the signal verification. A novel dynamic load altering attack that changes a group of unsecured loads is proposed in [21], which can cause physical damage to grid. A real-time anomaly detection algorithm that utilizes load forecasts, generation schedules and synchrophasor data can detect false data injection attack [22]. The vulnerabilities of standard load forecasting algorithms are examined in [23]. Though [21]–[23] investigated power grid cyber-security, vulnerability associated with DR signals has not been considered. Also, [20] solely focused on load altering attack which damages the grid by destabilizing the frequency away from its nominal range. But the studies on the impact of malicious cyber-attack targeted on transmission line power flow against DR signals were never conducted.

To bridge the gap presented above, this paper first studies the benefits of incorporating DR in SCED. The DR considered in this work incentivizes participating customers to shift demand from peak to non-peak hours. Following this, a false demand response signal and load measurement injection (FSMI) cyber-attack model is proposed in this paper which affects the EMS operation as shown in Fig. 1 (b). The proposed FSMI attack aims to lead to physical line overloads by manipulating DR signals and load measurements, which illustrates the cyber-vulnerability of demand response that is being widely adopted in real-time secure operations. The rest of this paper is organized as follows. Section II depicts the SCED and SCED-

DR formulation and Section III describes the proposed FSMI attack model. The test system with various load profiles is described in Section IV. Section V presents the simulation results and discusses the benefits and cyber-vulnerability of demand response in real-time operations of electric power systems. Finally, Section VI draws the conclusions.

## II. SCED AND SCED-DR FORMULATION

The goal of SCED/SCED-DR is the least-cost dispatch of committed generators for a short-term forecasted demand; in this work the look-ahead horizon is 1 hour with 15-minute moving intervals. SCED/SCED-DR models are based on the simplified DC power flow model and they are subject to security constraints of the generators and transmission networks. The SCED/SCED-DR considered in this paper is based on a moving window method; in other words, it looks ahead multiple intervals while only the solution associated with the first interval would be implemented.

The objective function for SCED-DR minimizes the cost of operations along with the penalties associated with shifting non-critical demand as a DR action for 15-minute, 30-minute and 45-minute, as shown in (1). In the case of SCED, the penalty cost associated with demand shifting is disregarded as DR is not implemented.

$$\text{Min:} \sum_{g,t}(c_g P_{g,t}) + \sum_{n,t}(c_{DR}^{15} DR_{n,t}^{15} + c_{DR}^{30} DR_{n,t}^{30} + c_{DR}^{45} DR_{n,t}^{45}) \quad (1)$$

Equation (2) shows that the generator outputs are restricted by their minimum and maximum limits. The network constraints are represented in (3)-(6). Here, (3) denotes the reference bus of the system. The line flows are calculated using the simplified DC power flow model in (4). The thermal limits of the lines are imposed in (5). Finally, constraint (6) ensures supply and demand are balanced for each node.

$$P_g^{min} \leq P_{g,t} \leq P_g^{max}, \forall g \epsilon G, t \epsilon T \quad (2)$$
$$\theta_{slack,t} = 0, \forall t \epsilon T \quad (3)$$
$$P_{k,t} - b_k(\theta_{n,t} - \theta_{m,t}) = 0, \forall k \epsilon K, t \epsilon T \quad (4)$$
$$-P_k^{max} \leq P_{k,t} \leq P_k^{max}, \forall k \epsilon K, t \epsilon T \quad (5)$$
$$\sum_{g \epsilon G(n)} P_{g,t} + \sum_{k \epsilon K^+(n)} P_{k,t} - \sum_{k \epsilon K^-(n)} P_{k,t} = d_{n,t}, \forall n \epsilon N, t \epsilon T \quad (6)$$

In the case of SCED-DR, (6) is replaced with (7)-(8). The DR actions imply that non-critical demand can be shifted by one, two or three periods. As shown in (7), the amount of deferrable demand should not exceed the DR participation $DR_n^{Max}$ at each node, which is set to 30% of the nodal load in this work. The updated nodal power balance constraint (8) considers the load deferred from previous intervals to current intervals

$$DR_{n,t}^{15} + DR_{n,t}^{30} + DR_{n,t}^{45} \leq DR_n^{Max}, \forall n \epsilon N, t \epsilon T \quad (7)$$
$$\sum_{g \epsilon G(n)} P_{g,t} + \sum_{k \epsilon K^+(n)} P_{k,t} - \sum_{k \epsilon K^-(n)} P_{k,t} = d_{n,t} - DR_{n,t}^{15} - DR_{n,t}^{30} - DR_{n,t}^{45} + DR_{n,t-1}^{15} + DR_{n,t-2}^{30} + DR_{n,t-3}^{45}, \forall n \epsilon N, t \epsilon T \quad (8)$$

In summary, the regular SCED is modelled through (1)-(6) while the SCED-DR is modelled through (1)-(5) and (7)-(8).

## III. FSMI ATTACK MODELS

DR typically requires timely communication to control the electricity demand within a very short response time. In the proposed FSMI model, it is assumed that the attacker has the knowledge of the system topology and parameters and has access to all load measurements. It is also assumed that the attacker has control of the real-time communication channels to intercept and modify the DR signals sent by operator as shown in Fig. 1 (b).

The solution only to the first interval of SCED-DR is implemented per Section II; and the proposed FSMI model targets at the first interval or current operating interval. The generation dispatch, $P_{g,1}$, and nodal phase angle, $\theta_{n,1}$, obtained from SCED-DR are represented as $P_{g,1}^*$ and $\theta_{n,1}^*$, respectively; they are fixed parameters in the proposed FSMI attack model. The false DR signals, $\widetilde{DR}_{n,1}$, are modeled to overload a target line. Note that shortly after the proposed attack and when users follow the false DR signals, some false load measurements are required to cover such an FSMI attack. The proposed FSMI attack model can be formulated as an optimization problem which focuses on the first period of the SCED-DR model described in Section II. Its objective function (9) is to maximize the physical line flow of a chosen target line $l$ in the attack region for the first SCED interval.

$$\text{Max: } sgn(P_{l,1}) \cdot \tilde{P}_{l,1} \quad (9)$$

$$\tilde{P}_{k,1} - b_k(\tilde{\theta}_{n,1} - \tilde{\theta}_{m,1}) = 0, \forall k \epsilon K \quad (10)$$
$$\sum_{g \epsilon G(n)} P_{g,1}^* + \sum_{k \epsilon K^+(n)} \tilde{P}_{k,1} - \sum_{k \epsilon K^-(n)} \tilde{P}_{k,1} = d_{n,1} - \widetilde{DR}_{n,1}, \forall n \epsilon N \quad (11)$$
$$DR_{n,1} - DR_{n,1}^{15*} - DR_{n,1}^{30*} - DR_{n,1}^{45*} = 0, \forall n \epsilon N \quad (12)$$
$$c_{n,1} = \theta_{n,1}^* - \tilde{\theta}_{n,1}, \forall n \epsilon N \quad (13)$$
$$\Delta \widetilde{DR}_{n,1} = DR_{n,1} - \widetilde{DR}_{n,1}, \forall n \epsilon N \quad (14)$$
$$\widetilde{DR}_{max} = (1 + \alpha) \cdot DR_{n,1}, \forall n \epsilon N \quad (15)$$
$$\widetilde{DR}_{min} = (1 - \alpha) \cdot DR_{n,1}, \forall n \epsilon N \quad (16)$$
$$\widetilde{DR}_{min} \leq \widetilde{DR}_{n,1} \leq \widetilde{DR}_{max}, \forall n \epsilon N \quad (17)$$
$$\sum_{n \epsilon N} |c_{n,1}| \leq S_0, \forall n \epsilon N \quad (18)$$
$$\sum_{n \epsilon N} |\Delta \widetilde{DR}_{n,1}| \leq Q_0, \forall n \epsilon N \quad (19)$$

The cyber attacked line power flows are obtained through (10) using the DC model and the cyber nodal power balance is enforced using (11). The total load that is expected to be shifted by SCED-DR without FSMI attack is calculated in (12). Here, the amount of demand deferred from the first period to the subsequent 3 periods, $DR_{n,1}^{15*}$, $DR_{n,1}^{30*}$ and $DR_{n,1}^{45*}$, are the fixed values obtained from SCED-DR. The relationship between physical bus angles and cyber bus angles is shown in (13). Equation (14) defines the demand response deviation, $\Delta \widetilde{DR}_{n,1}$, as the difference between scheduled demand response and attacked demand response. Constraints (15)-(17) limits the attack region to avoid large DR signal changes that can intimidate the operators with practical considerations: (i) the attacker can only alter DR signals for loads that are supposed to receive the actual DR signals, and (ii) DR signal deviation is limited to a tolerance set by the attacker. The summations of the absolute change in state variables and DR deviations are restricted by (18) and (19) respectively. The $l_l$-norm constraints (18) and (19) can be converted into equivalent linear constraints [15] which are represented in (20)-(25).

$$c_n \leq s_n, \forall n \epsilon N \quad (20)$$
$$-c_n \leq s_n, \forall n \epsilon N \quad (21)$$

$$\sum_n s_n \leq S_0, \forall n \epsilon N \quad (22)$$
$$\Delta \widetilde{DR}_{n,1} \leq q_n, \forall n \epsilon N \quad (23)$$
$$-\Delta \widetilde{DR}_{n,1} \leq q_n, \forall n \epsilon N \quad (24)$$
$$\sum_n q_n \leq Q_0, \forall n \epsilon N \quad (25)$$

A limited-channel FSMI attack model can then be formed using (9)-(19) and (20)-(25). However, the attacker can magnify the severity of the attack by sending erroneous signals to controllable loads that did not receive any DR signal from the operator in the pre-attack state. This type of attack is known as an unlimited-channel FSMI attack model which is represented by (9)-(16) and (20)-(26). Here, (26) relaxes the constraints on the number of attacked DR channels and the amount of nodal DR deviation that are restricted in (17).

$$\widetilde{DR}_{n,1} \leq DR_n^{Max}, \forall n \epsilon N \quad (26)$$

## IV. TEST CASE: IEEE 24-BUS SYSTEM WITH RES

To study the effect of DR, the IEEE 24-bus system [24], is utilized. The base system contains 24 buses, 33 generators and 38 lines. The total generation capacity from generators is 3,393 MW and the system peak load is 2,281 MW.

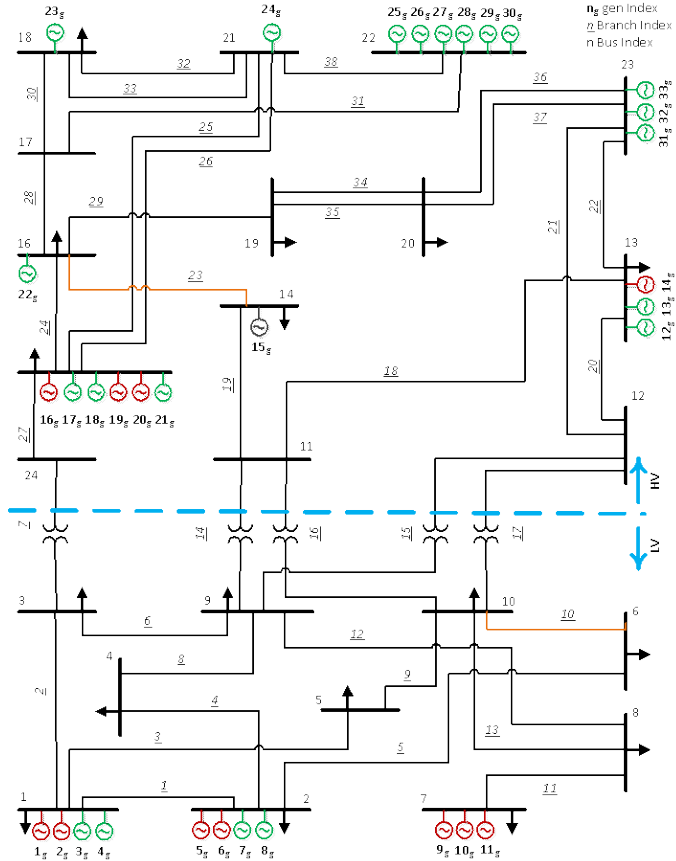

Fig. 2. IEEE 24-bus system with on-line and off-line generators [25].

Based on unit-commitment for the forecasted demand, only 21 generators are online. In Fig.2, online generators that have a combined capacity of 2,792 MW are shown in green and off-line generators are shown in red. Line 10 and line 23 are highlighted in orange as they are the bottle neck lines in the system. Fig. 3 shows the multiple one-hour system demand profiles with four 15-minute intervals. Three different scenarios are considered to identify the impact of DR: low, medium and high varying demand scenarios.

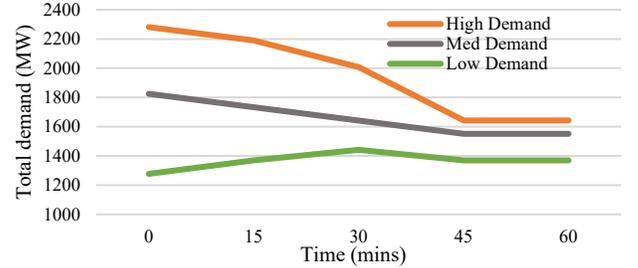

Fig. 3. System demand in various intervals.

## V. RESULTS AND ANALYSIS

The mathematical model was implemented using AMPL and solved using Gurobi with a MIPGAP of 0.001. The computer with Intel® Xeon(R) W-2195 CPU @ 2.30GHz, and 128 GB of RAM was utilized.

### A. Benefits of DR:

From Fig. 4, SCED-DR provides a lower cost solution compared to SCED for medium and high varying demand. At high demand variation, implementation of DR results in a total of 376 MW shifted by 15-minute, 140 MW shifted by 30-minute and 99 MW shifted by 45-minute which provides a savings of $2,986. For medium demand, the savings are about $1,100 and DR actions are lower compared to high demand scenario. For low demand scenario, the system is operating well below its capacity and as a result, both SCED and SCED-DR provides the same solution. Here, DR actions are not activated since none of the lines in the system are congested.

$$LF = \frac{\sum_{n,t}(netDmd_{n,t} * \Delta T)}{\sum_{n,t} netDmd_n^{max} \Delta T} * 100\% \quad (27)$$

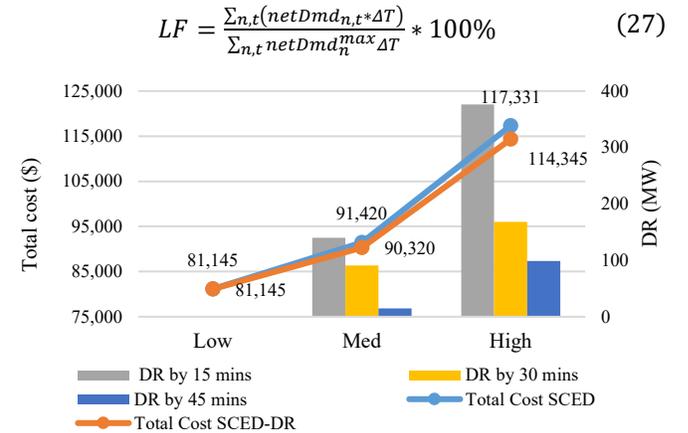

Fig. 4. Total system cost and cumulative DR shifted for different scenarios.

The implementation of DR also flattens load profile as shown in Fig. 5. This directly correlates as an increase in system short-term load factor utilization (LF). In this paper, the short-term LF is defined as the ratio of the average load to the peak load during the entire time period. This is represented in (27). The *LF* in SCED for medium and high demand scenario are 92.5% and 89% respectively. Whereas, in SCED-DR, the *LF* for medium and high demand scenario is 96.8% and 96.4% respectively. For low demand scenario, the LF is unchanged in SCED and SCED-DR as there are no effective DR actions required in SCED-DR. As DR actions takes place, it leads to higher load factors which indicates an increase of efficient utilization of system resources.

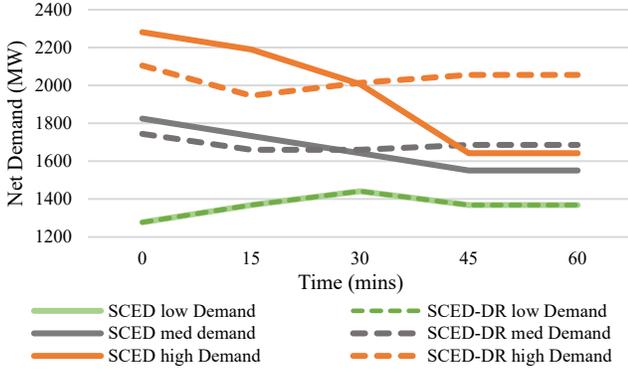

Fig. 5. Net demand served for various intervals in SCED/SCED-DR.

*B. FSMI Cyber-Attack:*

The line loading rate is defined as the ratio of line power flow to line thermal rating; the pre-attack line loading rates of all lines for the high demand scenario are obtained from SCED-DR and are represented in Fig. 6. The limited-channel FSMI attack model limits the attacker to focus on the channels that receive DR signals from operators. Therefore, to damage the system, the attacker should first select a targeted line to operate his attack. From Fig. 6 that shows the initial system condition, the plausible target lines are line 10, line 28 and line 23 which have the highest loading rates in the system. It can be noted that line 10 and line 28 have a loading rate less than 100%, which implies that these lines are uncongested in the pre-attack state; whereas, line 23 with a loading rate of 100% implies that the line is congested or carrying a flow at its capacity. For line 10, the power flow during pre-attack is 139.6 MW while the thermal rating of line 10 is 157.5 MW. For line 28, the power flow before attack is 400.7 MW and its thermal rating is 450 MW. For line 23, the power flow before attack is at its thermal rating limit, 315 MW.

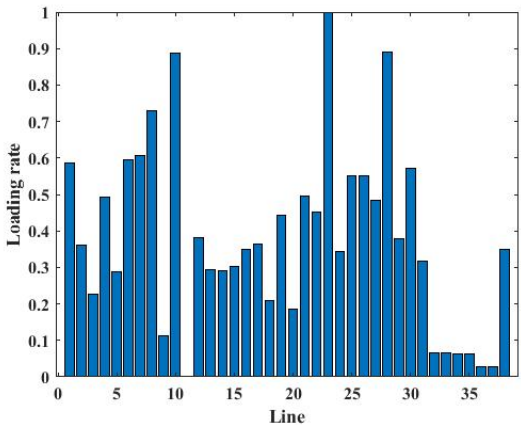

Fig. 6. Pre-attack line loading rate of all system lines.

To study the effectiveness of the proposed limited-channel FSMI cyber-attack, we highlight results for three targeted instances: (i) line 10 as targeted line, (ii) line 28 as targeted line and (iii) line 23 as targeted line. Sensitivity analysis is conducted to examine how the parameter $\alpha$ affects the performance of the proposed limited-channel FSMI attack. This is done by varying $\alpha$ from 10% to 100% which directly changes the limit of DR deviation in the FSMI attack model. Fig. 7 illustrates the line loading rate of those three lines under three attacks with target on line 10, 28 and 23 respectively. All three DR attacks result in an increase in actual target line power flow; the increase for the attack targeting on line 10 is the greatest. However, it is not overloaded even when the load shift deviation factor is set to 100%; this is because the loading rate of line 10 is low in the pre-attack state. For line 23, a congested line in the pre-attack state, can be easily and significantly overloaded. From the perspective of an attacker, an attack on a congested line would be very effective even with limited resources.

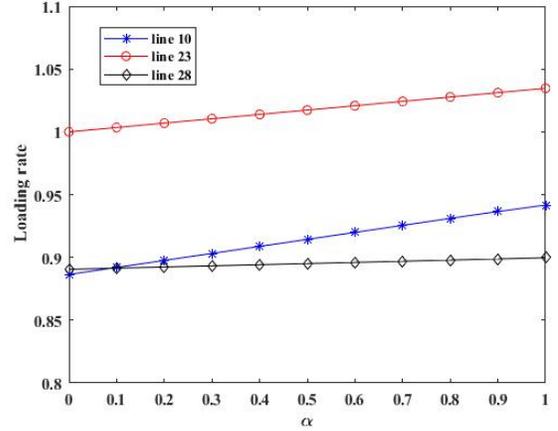

Fig. 7. Loading rate of line 10, line 28 and line 23 for a limited-channel FSMI attack with the attack target of line 10, line 28 and line 23 respectively.

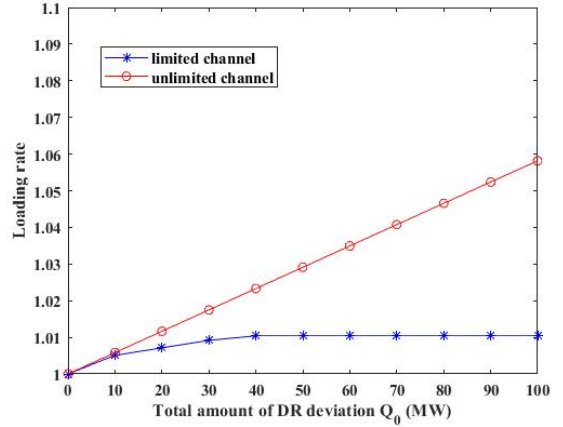

Fig. 8. Loading rate of line 23 with a limited-channel FSMI attack and an unlimited-channel FSMI attack model with the attack target of line 23.

Fig. 8 shows the changes of line 23 loading rates with various settings of the total amount of DR deviation $Q_0$ from 0 to 100 MW, under a limited-channel FSMI attack and an unlimited-channel FSMI attack respectively. The maximal power flow on targeted congested line 23 increases slightly and then remain stable with a maximum line overload of 3.3 MW for a limited-channel FSMI attack; however, the flow overload on the same line exceeds 18.9 MW when an unlimited-channel FSMI attack model is utilized. In Fig. 8, $\alpha$ is set 0.3.

The impact of an unlimited-channel FSMI attack model on various demand scenarios was studied using the three demand scenarios: low, medium and high demand scenarios, where DR deviation $Q_0$ is set 100 MW. The SCED-DR solution for high demand and medium demand emphasizes that line 23 is always congested, which implies it is more likely to be chosen as the targeted line. Fig. 9 illustrates the effect of the demand level in an unlimited-channel FSMI attack model on target branch 23. The simulation shows that loading rates always increase on the targeted line. Especially, under high demand and medium demand scenarios, the loading rate of targeted line increases considerably whereas the increase in targeted line loading rate is marginal for low demand scenario.

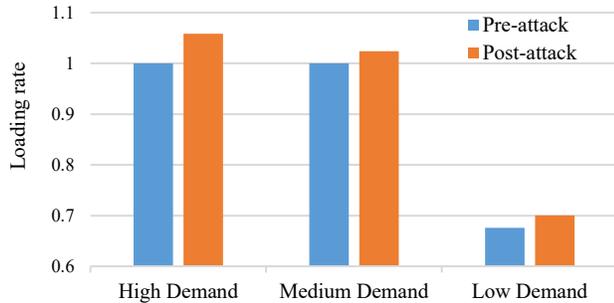

Fig. 9. Change in line loading rate on targeted line 23 using an unlimited-channel FSMI attack for different demand levels.

## VI. Conclusions

A SCED model incorporating demand shifting as DR action for real-time applications was presented. It can be noted that DR leads to substantial cost savings especially when the system is congested. DR also leads to significant flattening of load profile in the short term by shifting demands in peak periods to non-peak periods. SCED-DR under low demand resulted in 0 MW of demand shifted whereas medium and high demand resulted with 246 MW and 643 MW of cumulative DR actions respectively, for the entire time period. This corresponds directly to a higher short-term load factor usage of 96.8% and 96.4% in the case of medium and high demand scenarios, respectively. However, if these DR channels are hacked, the system can be vulnerable to targeted cyber-attacks that affect the system adversely.

The proposed FSMI cyber-attack model can effectively manipulate the DR signals to maximize the power flow on a targeted line and thus overload the targeted line. Initially, the effectiveness of the limited-channel FSMI attack was analyzed by modelling the attack on various targeted lines along with their sensitivity to the key parameter $\alpha$. Simulation results show that congested lines are more sensitive to the cyber-attack than non-congested lines and they can be overloaded with ease. Finally, the unlimited-channel FSMI attack was examined, and numerical simulations show that the unlimited-channel FSMI attack leads to much larger line overloads than the limited-channel FSMI attack. It is noted that high demand scenario may cause more significant overload than other scenarios.

## VII. References

[1] Xingpeng Li and K. W. Hedman, "Enhanced Energy Management System with Corrective Transmission Switching Strategy—Part I: Methodology" *IEEE Transactions on Power Systems*, vol. 34, no. 6, pp. 4490-4502, Nov. 2019.
[2] K. E. Van Horn, A. D. Domínguez-García and P. W. Sauer, "Measurement-Based Real-Time Security-Constrained Economic Dispatch," *IEEE Transactions on Power Systems*, vol. 31, no. 5, pp. 3548-3560, Sep. 2016.
[3] C. Su and D. Kirschen, "Quantifying the Effect of Demand Response on Electricity Markets," *IEEE Transactions on Power Systems*, vol. 24, no. 3, pp. 1199-1207, Aug. 2009.
[4] Demand response strategies, PJM [Online]. Available: https://www.pjm.com/markets-and-operations/demand-response.aspx
[5] Load Resource Participation in the ERCOT Markets, ERCOT [Online]. Available: http://www.ercot.com/services/programs/load/laar
[6] Demand response and load participation, CAISO [Online]. Available: http://www.caiso.com/participate/Pages/Load/Default.aspx
[7] Demand resources, ISONE [Online]. Available: https://www.iso-ne.com/markets-operations/markets/demand-resources/
[8] Schedule 30 FERC Electric Tariff Emergency Demand Response Initiative Schedules 36.0.0, MISO [Online]. Available: https://cdn.misoenergy.org/Schedule%2030109703.pdf
[9] Demand Response, NYISO [Online]. Available: https://www.nyiso.com/demand-response
[10] G.N. Ericsson, "Cyber Security and Power System Communication—Essential Parts of a Smart Grid Infrastructure," *IEEE Transactions on Power Delivery*, vol. 25, no. 3, pp. 1501-1507, Aug. 2010.
[11] H. Sandberg, A. Teixeira, and K. H. Johansson, "On security indices for state estimators in power networks," in Proc. *1st Workshop Secure Control Syst.*, 2010.
[12] J. Kim and L. Tong, "On topology attack of a smart grid: Undetectable attacks and countermeasures," *IEEE J. Select. Areas Commun.*, vol. 31, no. 7, pp. 1294–1305, Jul. 2013.
[13] D. Kundur, X. Feng, S. Liu, T. Zourntos, and K. Butler-Purry, "Towards a framework for cyber-attack impact analysis of the electric smart grid," in *Proc. 2010 1st IEEE Int. Conf. Smart Grid Communications (SmartGridComm)*, Oct. 2010, pp. 244–249.
[14] L. Jia, J. Kim, R. Thomas, and L. Tong, "Impact of data quality on real-time locational marginal price," *IEEE Trans. Power Syst.*, vol. 29, no. 2, pp. 627–636, Mar. 2014.
[15] J. Liang, L. Sankar and O. Kosut, "Vulnerability analysis and consequences of false data injection attack on power system state estimation," *IEEE Power & Energy Society General Meeting*, Chicago, IL, Jul. 2017.
[16] Xingpeng Li and K. W. Hedman, "Enhancing Power System Cyber-Security with Systematic Two-Stage Detection Strategy," *IEEE Transactions on Power Systems*, vol. 35, no. 2, pp. 1549-1561, Mar. 2020.
[17] Apurva Mohan and Daisuke Mashima,"Towards Secure Demand-Response Systems on the Cloud," *IEEE International Conference on Distributed Computing in Sensor Systems,* Marina Del Rey, CA, USA, Jun. 2014.
[18] 2019 Assessment of Demand Response and Advanced metering, FERC [Online]. Available: https://www.ferc.gov/legal/staff-reports/2019/DR-AM-Report2019.pdf
[19] Ye Yan, Yi Qian, Hamid Sharif and David Tipper, "A Survey on Cyber Security for Smart Grid Communications," *IEEE Communications Surveys & Tutorials*, vol. 14, no. 4, pp. 998-1010, Jan. 2012.
[20] Sajjad Amini, Fabio Pasqualetti and Hamed Mohsenian-Rad," Dynamic Load Altering Attacks Against Power System Stability: Attack Models and Protection Schemes," *IEEE Transactions on Smart Grid*, vol. 9, no. 4, pp. 2862-2872, Jul. 2018.
[21] Daisuke Mashima, Ulrich Herberg and Weipeng Chen, "Enhancing Demand Response signal verification in automated Demand Response systems," *Innovative Smart Grid Technologies*, Washington, DC, USA, May 2014.
[22] Aditya Ashok, Manimaran Govindarasu and Venkataramana Ajjarapu, "Online Detection of Stealthy False Data Injection Attacks in Power System State Estimation," *IEEE Transactions on Smart Grid*, vol. 9, no. 3, pp. 1636-1646, May. 2018
[23] Yize Chen, Yushi Tan and Baosen Zhang, "Exploiting Vulnerabilities of Load Forecasting Through Adversarial Attacks," *Proceedings of the Tenth ACM International Conference on Future Energy Systems*, Phoenix AZ USA, June. 2019
[24] C. Grigg *et al*., "The IEEE Reliability Test System-1996. A report prepared by the Reliability Test System Task Force of the Application of Probability Methods Subcommittee," *IEEE Transactions on Power Systems*, vol. 14, no. 3, pp. 1010-1020, Aug. 1999.
[25] Arun Venkatesh Ramesh and Xingpeng Li, "Security Constrained Unit Commitment with Corrective Transmission Switching," *North American Power Symposium (NAPS)*, Wichita, KS, USA, Oct. 2019.